\documentclass[pra,twocolumn,floatfix,showpacs,showkeys,preprintnumbers]{revtex4}
\usepackage{amssymb}
\usepackage{amsmath}
\usepackage{graphicx}% Include figure files
\usepackage{dcolumn}% Align table columns on decimal point
\usepackage{bm}% bold math
\usepackage[mathscr]{eucal}
\usepackage{epsfig}
\usepackage{rotating}

\begin{document}
%\preprint{ }
\title{Realization of Arbitrary Gates in Holonomic Quantum Computation}% Force line breaks with \\
\author{Antti\ O.\ Niskanen}
%\email{aniskane@focus.hut.fi}
\author{Mikio\ Nakahara} 
\altaffiliation[Also at ]{Department of Physics, Kinki University, Higashi-Osaka 577-8502, Japan.}
%\email{nakahara@math.kindai.ac.jp}
\author{Martti\ M.\ Salomaa}
%%\email{martti.salomaa@hut.fi}
\affiliation{Materials Physics Laboratory, POB 2200 (Technical Physics), FIN-02015 HUT, 
Helsinki University of Technology, Finland}

\begin{abstract}
Among the many proposals for the realization of a quantum computer,
holonomic quantum computation (HQC) is distinguished from
the rest in that it is geometrical in nature and thus expected to be
robust against decoherence.
Here we analyze the realization of various quantum gates
by solving the inverse problem: Given a unitary matrix,
we develop a formalism by which we find loops in the parameter space
generating this matrix as a holonomy. We demonstrate for the first time that such a
one-qubit gate as the Hadamard gate and such two-qubit gates
as the CNOT gate, the SWAP gate and the discrete Fourier transformation
can be obtained with a single loop.

\end{abstract}

%Use showkeys class option if keyword
\maketitle

\section{Introduction}\label{sec1}

Quantum computing is an emerging scientific discipline, 
in which the merging and mutual cross-fertilization of two of the most important developments 
in physical science and information technology of the past century --- quantum mechanics and computing --- 
has resulted in an extraordinarily rapid rate of progress of interdisciplinary nature.  
Interesting problems to address in this context include fundamental questions as to what are the ultimate 
physical limits of computation and communication. For introductions to quantum computing and quantum 
information processing see, e.g., Refs.~\cite{review,book,gruska}.

Holonomic quantum computation (HQC) was first suggested by Zanardi and Rasetti in Ref.~\cite{holonomic}.
The concept has been further developed in Refs.~\cite{fujii,univ,nonabelian,holonomies,optical}.
The suggestion is very intriguing itself; quantum-logical operations are achieved by
driving a degenerate system around adiabatic loops in the parameter manifold. The resulting
gates are a generalization of the celebrated Berry phase \cite{berry} to encompass a degenerate
system. These are, in fact, non-Abelian holonomies. Due to
the geometric nature of these gates, quantum information processing is expected to be fault tolerant.
For instance the issue of timing and the lack of spontaneous decay are definite strengths of HQC.
Here we study the construction of holonomic quantum-logic gates numerically for the first time
via solving a certain inverse problem. Namely, we find the loop $\hat{\gamma}$ corresponding
to the desired unitary operator $\hat{U}$ by solving a high-dimensional optimization task.

The paper is organized as follows: In Section~\ref{sec2}, we present the physical and mathematical background
underlying our approach.
Sections~\ref{sec3},~\ref{sec4} and~\ref{sec5} comprise
the main part of the present paper. Loop parameterizations for one-
and two-qubit gates are presented in Section~\ref{sec3}.
The numerical method is introduced in Section~\ref{sec4}.
Then the optimal realization of a unitary gate as a holonomy
associated with a loop in the parameter space is investigated numerically
in Section~\ref{sec5}. Section~\ref{sec6} discusses the results.

\section{Hamiltonian and Holonomy}\label{sec2}

Here we first review the concept of
non-Abelian
holonomy to establish notation conventions.
Let us consider a family of Hamiltonians $\{H_{\lambda}\}$.
The point $\lambda$, continuously parameterizing the Hamiltonian,
is an element of a
manifold $\mathcal{M}$ called the control manifold and
the local coordinate of $\lambda$ is denoted by
$\lambda^i\ (1 \leq i \leq m=\mathrm{dim} \mathcal{M})$.
It is assumed that there exists only a finite number of eigenvalues
$\varepsilon_k(\lambda)\ (1 \leq k \leq R)$ for an arbitrary $\lambda
\in \mathcal{M}$ and that no level crossings occur. 
Suppose the $n$-th eigenvalue $\varepsilon_n(\lambda)$ is $g_n$-fold
degenerate for any $\lambda \in \mathcal{M}$ and $\sum_{n=1}^R
g_n =N$. The degenerate subspace at $\lambda$
is denoted by $\mathcal{H}_n(\lambda)$. Accordingly, the Hamiltonian is
expressed as an $N \times N$ 
%block-diagonal 
matrix.
%with $R$ blocks,
%each block being expressed as a $g_n \times g_n$ matrix.
The orthonormal basis vectors of $\mathcal{H}_n(\lambda)$ are
denoted by $\{| n \alpha ;\lambda \rangle\}$;
$$
H_{\lambda}|n \alpha; \lambda \rangle = \varepsilon_n(\lambda)
|n \alpha; \lambda \rangle,\quad
\langle n \alpha; \lambda | m \beta ; \lambda \rangle = \delta_{
mn} \delta_{\alpha \beta}.
$$
Note that there are $U(g_n)$ degrees of freedom in the choice of the basis
vectors $\{ |n\alpha; \lambda\rangle\}$.

Let us now assume that the parameter $\lambda$ is changed adiabatically.
We will be concerned with a particular subspace, say the
ground state $\mathcal{H}_1(\lambda)$ and we
drop the index $n$ to simplify the notation. Suppose the initial state
at $t=0$ is an eigenstate $|\psi_{\alpha}(0) \rangle = |\alpha;
\lambda(0) \rangle$ with the energy $\varepsilon=0$ possibly through shifting
the zero-point of the energy. In fact, we are not interested in the
dynamical phase at all
and hence assume that the eigenvalue in this subspace vanishes
for any $\lambda \in \mathcal{M}$. 
The Schr\"odinger equation is
\begin{equation}
i \frac{d}{dt} | \psi_{\alpha} (t) \rangle = H_{\lambda(t)}|\psi_{\alpha}
(t) \rangle,
\label{eq:2.1}
\end{equation}
whose solution may be assumed to take the form
\begin{equation}
|\psi_{\alpha}(t) \rangle = \sum_{\beta=1}^{g} |\beta; \lambda(t) \rangle
U_{\beta \alpha}(t).
\label{eq:2.2}
\end{equation}
The unitarity of the matrix $U_{\beta \alpha}(t)$ follows from the
normalization of $|\psi_{\alpha} (t) \rangle$. By substituting
Eq.~(\ref{eq:2.2}) into (\ref{eq:2.1}), one finds that $U_{\beta \alpha}$ satisfies
\begin{equation}
\dot{U}_{\beta \alpha}(t) = -\sum_{\gamma} \left\langle \beta; \lambda(t)\left|
\frac{d}{dt} \right|\gamma; \lambda(t) \right\rangle U_{\gamma \alpha}.
\end{equation}
The formal solution may be expressed as
\begin{eqnarray}
U(t)&=& \mathcal{T} \exp\left(- \int_0^t A(\tau) d\tau \right)\nonumber\\
&=& I -\int_0^t A(\tau) d\tau \nonumber\\
& & +\int_0^t d\tau \int_0^\tau d\tau'
A(\tau) A(\tau') + \ldots,
\end{eqnarray}
where $\mathcal{T}$ is the time-ordering operator and
$$
A_{\beta \alpha}(t) = \left\langle \beta; \lambda(t)\left|
\frac{d}{dt} \right|\alpha; \lambda(t) \right\rangle.
$$
Let us introduce the %$\mathfrak{u}(g)$
Lie-algebra-valued connection
\begin{equation}
\mathcal{A}_{i, \beta \alpha} =
\left\langle \beta; \lambda(t)\left|
\frac{\partial}{\partial\lambda^i} \right|\alpha; \lambda(t) \right\rangle
\end{equation}
through which $U(t)$ is expressed as
\begin{equation}
U(t) = \mathcal{P} \exp\left( - \int_{\lambda(0)}^{\lambda(t)} 
\mathcal{A}_{i}d \lambda^i\right),
\end{equation}
where $\mathcal{P}$ is the path-ordering operator. Note that $\mathcal{A}_i$
is anti-Hermitian, $\mathcal{A}_i^{\dagger} = - \mathcal{A}_i$.

Suppose the path $\lambda(t)$ is a loop $\gamma(t)$ in $\mathcal{M}$ such that
$\gamma(0) = \gamma(T) = \lambda_0$. Then it is found after traversing $\gamma$
that one ends up with the state
\begin{equation}
|\psi_{\alpha}(T) \rangle = \sum_{\beta=1}^g |\psi_{\beta}(0) \rangle
U_{\beta \alpha}(T)
\end{equation}
where use has been made of the definition $|\psi_{\beta}(0) \rangle
= |\beta; \lambda_0 \rangle$. The unitary matrix
\begin{equation}\label{eq:U}
U_\gamma\equiv U(T) =
\mathcal{P} \exp \left(-\oint_{\gamma}
\mathcal{A}_i d \gamma^i \right)
\end{equation}
is called the holonomy associated with the loop $\gamma(t)$.
Note that $U_\gamma$ is independent of the parameterization of
the path but only depends upon its geometric image in $\mathcal{M}$.

The space of all the loops based at $\lambda_0$ is denoted
\begin{equation}
L_{\lambda_0}(\mathcal{M}) = \{ \gamma:[0, T] \to \mathcal{M}|
\gamma(0) = \gamma(T) = \lambda_0\}.
\end{equation}
The set of the holonomy
\begin{equation}
{\mathrm{Hol}}(\mathcal{A}) =\{ U_\gamma|\gamma \in L_{\lambda_0}(\mathcal{M}) \}
\end{equation}
has a group structure \cite{nakahara} and is called the holonomy group. 
It is clear that $\mathrm{Hol}(\mathcal{A}) \subset U(g)$. The
connection $\mathcal{A}$ is called irreducible when ${\mathrm{Hol}}(
\mathcal{A}) = U(g)$. 

\section{Three-State Model and Quantum-Gate Construction}\label{sec3}

\subsection{One-qubit gates}

To make things tractable, we employ a
simple model Hamiltonian called the three-state model as the basic building
block for our strategy. This is a 3-dimensional Hamiltonian with the matrix
form
\begin{equation}\label{eq:11}
H_{\lambda_0} = \epsilon |2 \rangle \langle 2|
= \left( \begin{array}{ccc}
\epsilon&0&0\\
0&0&0\\
0&0&0
\end{array} \right).
\end{equation}
The first column (row) of the matrix refers to the auxiliary state
$|2 \rangle$ with the energy $\epsilon>0$
while the second and the third columns (rows) refer to the vectors
$|0 \rangle$ and $|1 \rangle$, respectively, with vanishing energy.
The qubit consists of the last two vectors.

The control manifold of the Hamiltonian~(\ref{eq:11}) is 
the complex projective space $\mathbb{C}P^2$. This is seen most directly 
as follows: The most general form of the iso-spectral deformation
of the Hamiltonian is of the form $H_{\gamma} \equiv
W_{\gamma} H_{\lambda_0} W_{\gamma}^{\dagger}$,
where $W_{\gamma} \in U(3)$. Note, however, that not all the elements
of $U(3)$ are independent. It is clear that $H_{\gamma}$ is independent of
the overall phase of $W_{\gamma}$, which reduces the number of degrees of freedom from $U(3)$
to $U(3)/U(1)=SU(3)$. Moreover, any element of $SU(3)$ may be decomposed into
a product of three $SU(2)$ matrices as follows
\begin{equation}
W_{\gamma}=\underbrace{\left(\begin{matrix}\bar{\beta}_1&\bar{\alpha}_1&0\\ 
 -\alpha_1&\beta_1&0\\0&0&1
\end{matrix}\right)}_{U_1}
\underbrace{\left(\begin{matrix}\bar{\beta}_2&0&\bar{\alpha}_2\\ 
0&1&0\\
 -\alpha_2&0&\beta_2
\end{matrix}\right)}_{U_2}
\underbrace{\left(\begin{matrix}1&0&0\\0&\bar{\beta}_3&\bar{\alpha}_3\\ 
 0&-\alpha_3&\beta_3
\end{matrix}\right)}_{U_3},
\end{equation}
which is know as the Givens decomposition. Here
$\alpha_j = e^{i \phi_j} \sin \theta_j$ and $\beta_j = e^{i \psi_j}
\cos \theta_j$. It is clear that
$H_{\gamma}$ is independent of $U_3$ since $[H_{\lambda_0} , U_3] = 0$.
This further reduces the physical degrees of freedom to $SU(3)/SU(2) \cong S^5$.
The product $U_1 U_2$ contains six parameters while $S^5$ is five-dimensional;
there must be one redundant parameter in $U_1 U_2$. This parameter is easily
found out by writing the product explicitly. The result depends
only on the combination $\phi_2 - \psi_2$ and not on individual parameters.
Accordingly, we may redefine $\phi_2$ as $\phi_2 - \psi_2$ to eliminate
$\psi_2$. Furthermore, after this redefinition we find that the Hamiltonian
depends only on $\phi_1-\psi_1$ and $\phi_2-\psi_1$ and hence $\psi_1$
may also be subsumed
by redefining $\phi_1$ and $\phi_2$, which reduces
the independent degrees of freedom down to $\mathbb{C}P^2 \cong S^5/S^1$.

Let $[z^1, z^2, z^3]$
be the homogeneous coordinate of $\mathbb{C}P^2$ and $(1, \xi_1, \xi_2)$
be the corresponding inhomogeneous coordinate, where $\xi_1 = z^2/z^1,
\xi^2 = z^3/z^1$ in the coordinate neighborhood with $z^1 \neq 0$.
If we write $\xi_k = r_k e^{i \varphi_k}$, the above correspondence, i.e. the
embedding of $\mathbb{C}P^2$ into $U(3)$, is
explicitly given by $\theta_k = \tan^{-1} r_k$ and $\phi_k = \varphi_k$. 

The connection coefficients are easily calculated in the present model
and are given by
\begin{equation}\label{eq:Atheta1}
\mathcal{A}_{\theta_1}=
\left ( \begin {matrix}  0&-\sin\theta_2 e^{-i\left ( \phi_2-\phi_1\right )}\\
\sin {\theta_2}e^{i\left (\phi_2-\phi_1\right )}&0\end {matrix}\right )\,,
\end{equation}
\begin{equation}\label{eq:Atheta2}
\mathcal{A}_{\theta_2}=\left(\begin{matrix}0&0\\0&0
\end{matrix}\right)\,,
\end{equation}
\begin{widetext}
\begin{equation}\label{eq:Aphi1}
\mathcal{A}_{\phi_1}=\left(\begin{matrix}
-i\sin^2\theta_1&-\frac{i}{2}\sin2\theta_1\sin\theta_2{e^{i(\phi_1-\phi_2)}}\\
-\frac{i}{2}\sin2\theta_1\sin\theta_2{e^{i(\phi_2-\phi_1)}}&i\sin^2\theta_2 \sin^2\theta_1
\end{matrix}\right)\,,
\end{equation}
\end{widetext}
%\begin{equation}\label{eq:Aphi1}
%A_{\phi_1}=\left(\begin{matrix}
%-i\sin^2\theta_1&-i\cos\theta_1\sin\theta_1\sin\theta_2{e^{i(\phi_1-\phi_2)}}\\
%-i\sin\theta_1\sin\theta_2\cos\theta_1{e^{i(\phi_2-\phi_1)}}&i\sin^2\theta_2 \sin^2\theta_1
%\end{matrix}\right)\,,
%\end{equation}
\begin{equation}\label{eq:Aphi2} \mathcal{A}_{\phi_2}=\left(\begin{matrix}
0&0\\0&-i\sin^2\theta_2
\end{matrix}\right)\,,
\end{equation}
where the first column (row) refers to $|0 \rangle$ while the second one refers to 
$|1 \rangle$. Using these connection coefficients, it is possible
to evaluate the holonomy associated with a loop $\gamma$ as
\begin{eqnarray}
U_\gamma&&= \\
&&\!\!\!\!\!\!\!\!\!\!\!\!\!\!\!
\mathcal{P} \exp\left(-\oint_{\gamma}
(\mathcal{A}_{\theta_1} d\theta_1 + \mathcal{A}_{\theta_2} d\theta_2 
+ \mathcal{A}_{\phi_1} d\phi_1 + 
\mathcal{A}_{\phi_2} d\phi_2)\right). \notag
\end{eqnarray}
Now our task is to find a loop that yields a given unitary matrix as
its holonomy.

\subsection{Two-qubit gates}

Let us consider a two-qubit reference Hamiltonian
\begin{equation}
H_{\lambda_0}^{\text{2-qubit}} = H_{\lambda_0}^a \otimes I_3 +
I_3 \otimes H_{\lambda_0}^b,
\end{equation}
where $H_{\lambda}^{a, b}$ are three-state Hamiltonians and $I_3$ is the
$3 \times 3$ unit matrix. Generalization to an arbitrary $N$-qubit
system is obvious. The Hamiltonian scales as $3^N$, instead of the $2^N$ in
the present model. It is also possible to consider a model with
$g$-degenerate eigenstates with one auxiliary state having finite
energy. This model, however, has a difficulty in realizing an entangled
state, without which the full computational power of a quantum
computer is impossible. 

We want to maintain the multipartite structure of the system
in constructing the holonomy. For this purpose, we separate
the unitary transformation into a product of single-qubit transformations
$(W_{\gamma}^a \otimes W_{\gamma}^b)$
and a purely two-qubit rotation
$W_{\gamma}^{\text{2-qubit}}$ which cannot be reduced into a tensor product
of single-qubit transformations. Therefore, we write the
iso-spectral deformation for a given loop $\gamma$ as
\begin{eqnarray}
H_{\gamma}^{\text{2-qubit}} &=& 
W_{\gamma}^{\text{2-qubit}} (W_{\gamma}^a \otimes W_{\gamma}^b)
H_{\lambda_0}^{\text{2-qubit}}\nonumber\\
& & \times (W_{\gamma}^a \otimes W_{\gamma}^b)^{\dagger}
W_{\gamma}^{\text{2-qubit} \dagger}.
\end{eqnarray}
The advantage of expressing the unitary matrix in this form is easily
verified when we write down the connection coefficients
for the one-qubit coordinates. Namely,
the two-qubit transformation does not affect the one-qubit transformation
at all;
\begin{eqnarray*}
\mathcal{A}_{i,\alpha \beta} &=& \left\langle \alpha; \lambda\left|
W_{\gamma}^{\dagger} \frac{\partial}{\partial \gamma^i} W_{\gamma} \right|\beta; \lambda
\right\rangle
\\
&=& 
\left\langle \alpha; \lambda \left|
(W_{\gamma}^a \otimes W_{\gamma}^b)^{\dagger} \frac{\partial}{\partial \gamma^i}
(W_{\gamma}^a \otimes W_{\gamma}^b) \right|\beta; \lambda \right\rangle,
\end{eqnarray*}
where $\gamma^i$ denotes a one-qubit coordinate.

There is a large number of possible choices for $W_{\gamma}^{\text{2-qubit}}$,
depending on the physical realization of the present scenario. 
To keep our analysis as concrete as possible, we have made the simplest
choice
\begin{equation}
W_{\gamma}^{\text{2-qubit}}  = W_{\xi} \equiv
e^{i \xi |11 \rangle \langle 11|}
\end{equation}
for our two-qubit unitary rotation. Let
\begin{widetext}
\begin{eqnarray*}
H_{\gamma}' &=& H_{\gamma}^a \otimes I_3 + I_3 \otimes H_{\gamma}^b\nonumber\\
&=& \left(\begin{smallmatrix}
h^a_{11}+h^b_{11}&h^b_{12}&h^b_{13}&h^a_{12}&0&0&h^a_{13}&0&0\\
h^b_{21}&h^a_{11}+h^b_{22}&h^b_{23}&0&h^a_{12}&0&0&h^a_{13}&0\\
h^b_{31}&h^b_{32}&h^a_{11}+h^b_{33}&0&0&h^a_{12}&0&0&h^b_{13}\\
h^a_{21}&0&0&h^a_{22}+h^b_{11}&h^b_{12}&h^b_{13}&h^a_{23}&0&0\\
0&h^a_{21}&0&h^b_{21}&h^a_{22}+h^b_{22}&h^b_{23}&0&h^a_{23}&0\\
0&0&h^a_{21}&h^b_{31}&h^b_{32}&h^a_{22}+h^b_{33}&0&0&h^a_{23}\\
h^a_{31}&0&0&h^a_{32}&0&0&h^a_{33}+h^b_{11}&h^b_{12}&h^b_{13}\\
0&h^a_{31}&0&0&h^a_{32}&0&h^b_{21}&h^a_{33}+h^b_{22}&h^b_{23}\\
0&0&h^a_{31}&0&0&h^a_{32}&h^b_{31}&h^b_{32}&h^a_{33}+h^b_{33}
\end{smallmatrix}\right).
\end{eqnarray*}
be a two-qubit Hamiltonian before $W_{\xi}$ is applied. Then after
the application of $W_{\xi}$ to $H_{\gamma}'$ we have the full Hamiltonian
\begin{eqnarray}
H_{\gamma}^{\text{2-qubit}} &=& W_{\xi} H_{\gamma}' W_{\xi}^{\dagger}
\nonumber\\
&=& \left(\begin{smallmatrix}
h^a_{11}+h^b_{11}&h^b_{12}&h^b_{13}&h^a_{12}&0&0&h^a_{13}&0&0\\
h^b_{21}&h^a_{11}+h^b_{22}&h^b_{23}&0&h^a_{12}&0&0&h^a_{13}&0\\
h^b_{31}&h^b_{32}&h^a_{11}+h^b_{33}&0&0&h^a_{12}&0&0&h^b_{13}e^ {-i \xi}\\
h^a_{21}&0&0&h^a_{22}+h^b_{11}&h^b_{12}&h^b_{13}&h^a_{23}&0&0\\
0&h^a_{21}&0&h^b_{21}&h^a_{22}+h^b_{22}&h^b_{23}&0&h^a_{23}&0\\
0&0&h^a_{21}&h^b_{31}&h^b_{32}&h^a_{22}+h^b_{33}&0&0&h^a_{23}e^ {-i \xi}\\
h^a_{31}&0&0&h^a_{32}&0&0&h^a_{33}+h^b_{11}&h^b_{12}&h^b_{13}e^ {-i \xi}\\
0&h^a_{31}&0&0&h^a_{32}&0&h^b_{21}&h^a_{33}+h^b_{22}&h^b_{23}e^ {-i \xi}\\
0&0&h^a_{31}e^ {i \xi}&0&0&h^a_{32}e^ {i \xi}&h^b_{31}e^ {i \xi}&h^b_{32}e^ {i \xi}&h^a_{33}+h^b_{33}
\end{smallmatrix}\right).
\end{eqnarray}
\end{widetext}

As for the connection, we find
\begin{equation}
\mathcal{A}_{\xi} = \left(\begin{matrix}
%0&0&0&0\\0&0&0&0\\0&0&0&0\\0&0&0&i\cos\theta^a_2\cos\theta^b_2
0&0&0&0\\0&0&0&0\\0&0&0&0\\0&0&0&i\cos^2\theta^a_2\cos^2\theta^b_2
\end{matrix}\right)
\end{equation}
where the columns and rows are ordered with respect to the basis
$\{|00 \rangle,|01 \rangle,|10 \rangle,|11 \rangle\}$.
It should be apparent from the above analysis that we can construct
an arbitrary controlled phase-shift gate with the help of a loop in the
$(\theta^2_a, \xi)$- or $(\theta^2_b, \xi)$-space. Accordingly, this gives the CNOT gate
with one-qubit operations, as shown below.

\subsection{Some Examples}

Before we proceed to present the numerical prescription to construct
arbitrary one- and two-qubit gates in the next section, it is instructive
to first work out some important examples whose loop can be constructed analytically. 
In particular, we will show that all the gates required
for the proof of universality may be obtained within the present
three-state model. 

The first example is the $\pi/8$-gate,
\begin{equation}
U_{\pi/8} =  \left( \begin{array}{cc}
1&0\\
0&e^{i \pi/8}
\end{array} \right).
\end{equation}
By inspecting the connection coefficients in Eqs.~(\ref{eq:Atheta1}-\ref{eq:Aphi2}), we easily
find that the loop presented by the sequence
\begin{eqnarray}
(\theta_2, \phi_2)&:&(0,0) \to (\pi/2,0) \to (\pi/2, \pi/8)\nonumber\\
& &\to (0, \pi/8) \to (0,0).
\end{eqnarray}
yields the desired gate.
Note that the loop is in the $(\theta_2, \phi_2)$-plane and that all the other
parameters are fixed at zero. Explicitly, we verify that
\begin{eqnarray}
U_{\pi/8} &=& \exp \left(
\frac{\pi}{8}\left. \mathcal{A}_{\phi_2} \right|_{\theta_2=0}\right)
\exp \left(
\frac{\pi}{2}\left. \mathcal{A}_{\theta_2} \right|_{\phi_2=\pi/8}\right)
\nonumber\\
& & \times \exp \left(-
\frac{\pi}{8}\left. \mathcal{A}_{\phi_2} \right|_{\theta_2=\pi/2}\right)
\exp \left(
-\frac{\pi}{2}\left. \mathcal{A}_{\theta_2} \right|_{\phi_2=0}\right)
\nonumber\\
&=& \exp \left(
-\frac{\pi}{8}\left. \mathcal{A}_{\phi_2} \right|_{\theta_2=\pi/2}\right).
\end{eqnarray}

The next example is the Hadamard gate
\begin{equation}
H = \frac{1}{\sqrt{2}} \left( \begin{array}{cc}
1&1\\
1&-1
\end{array} \right).
\end{equation}
Instead of constructing $H$ directly, we will rather use the decomposition
$$
H = e^{-i \pi/2} \exp\left(i \frac{\pi}{2} \sigma_z \right)
\exp\left(i \frac{\pi}{4} \sigma_y \right).
$$
It is easy to verify that the holonomy associated with the loop
\begin{eqnarray}
(\theta_2, \theta_1)&:&(0,0) \to (\pi/2,0) \to (\pi/2, \beta)\nonumber\\
& &\to (0, \beta) \to (0,0)
\end{eqnarray}
is $\exp(i \beta \sigma_y)$, while that associated with the loop
\begin{eqnarray}
(\theta_1, \theta_2, \phi_1)&:&(0,0, 0) \to (\pi/2,0, 0) \to (\pi/2, \pi/2,0)
\nonumber\\
& &\to (\pi/2, \pi/2, \alpha) \to (\pi/2,0, \alpha)
\nonumber\\
& &\to (0,0,\alpha) \to
(0,0,0)
\end{eqnarray}
is $\exp(i \alpha \sigma_z)$. Here again, the rest of the parameters are
fixed at zero. Finally, we construct the phase-shift gate $e^{i \delta }$,
which is produced by a sequence of two loops.
First we construct a gate similar to the $\delta$-shift gate using
(cf., the $\pi/8$-shift gate)
\begin{eqnarray}
(\theta_1, \phi_1)&:&(0,0) \to (\pi/2,0) \to (\pi/2, \delta)\nonumber\\
& &\to (0, \delta) \to (0,0).
\end{eqnarray}
This loop followed by the similar loop in the $(\theta_2, \phi_2)$-space
yields the $e^{i \delta}$-gate as
\begin{eqnarray}
&(\theta_1, \phi_1, \theta_2, \phi_2): \notag\\ 
&(0,0, 0,0) \to (0,0,\pi/2, 0) \to (0,0, \pi/2,\delta) \nonumber\\
& \to (0,0,0, \delta) \to (0,0,0, 0) \to (\pi/2, 0,0,0)
\nonumber\\
&\to (\pi/2,\delta, 0,0) \to (0, \delta,0,0) \to (0,0,0,0).
\end{eqnarray}

Finally, the controlled-phase gate $U(\Theta)=\exp(i\Theta|11\rangle\langle11|)$
can be written as
\begin{eqnarray}
(\theta_2^a, \xi)&:&(0,0) \to (\pi/2,0) \to (\pi/2, \Theta)\nonumber\\
& &\to (0, \Theta) \to (0,0).
\end{eqnarray}

\begin{figure}
\centering
\epsfig{figure=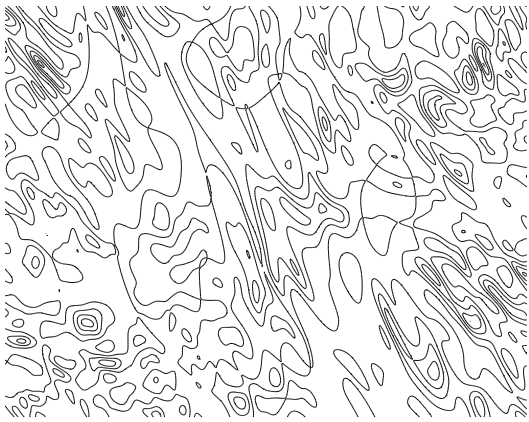, width=70mm}
\caption{\label{fg:scape}Objective function landscape in 2D.}
\end{figure}

\section{Numerical Method}\label{sec4}

Now we adopt a systematic approach to actually constructing arbitrary quantum gates. This is the first time
that arbitrary one- and two-qubit gates are constructed in a three-state model that is in a way the 
simplest possible realization for HQC while still maintaining the tensor-product structure necessary for exponential
speed-up. It has not been shown previously how to construct the CNOT, let alone the two-qubit Fourier
transform in a single loop. Hence, we resort to numerical methods. Since it is extremely difficult to see which single loop
results in a given unitary operator, our approach will be that of 
variational calculus. 

We convert the inverse
problem, i.e. which loop corresponds to a given unitary operator, to an optimization problem. The
problem of finding the unitary operator for a given loop is straightforward.
Keeping the basepoint of the holonomy loop fixed, we let the midpoints vary.
Owing to the $2\pi$-periodicity, the loops can end either in the origin or at any point 
that is 
modulo $2\pi$. 

The space of all possible loops is denoted by $\mathcal{V}$.  We shall restrict the variational task to the space of
polygonal paths $\mathcal{V}_k$, where $k$ is the number of vertices in the path excluding the basepoint.
Naturally, we have $\mathcal{V}_k\subset\mathcal{V}$ such that we are not guaranteed to find the
best possible solution among all the loops, but provided that we use a good optimization method, we may expect
to find the best solution in the limited space $\mathcal{V}_k$. Since the dimension of the variational space increases with $k$, one
is forced to use as low a $k$ as possible. For instance, for one-qubit gates the dimension is $4k$. 
In the case of two-qubit gates the dimension is $9k$. Low $k$ appears to be desirable for experimental reasons as well.

Formally, the optimization problem is to find a $\tilde{\gamma}$, such that 
\begin{equation}
f(\gamma)=\|\hat{U}-U_\gamma\|_F
\end{equation}
is minimized over all $\gamma\in\mathcal{V}_k$. 
We naturally hope the minimum value to be zero. 
Here $\|\cdot\|_F$ is the so-called Frobenius trace norm defined by
$\|{A}\|_F=\sqrt{\mathrm{Tr}\left({A}^\dagger{A}\right)}$.
We could employ the well-known conjugate-gradient method to solve the task at hand 
but this method, or any other derivative-based method, is not expected to perform well in the present 
problem due to the complicated structure of the objective function. 
Hence we will use the robust polytope algorithm \cite{poly}.

We have plotted a sample 2D section of the optimization space
in Fig.~\ref{fg:scape}. The axes represent two orthogonal directions
in the optimization space of a certain two-qubit gate. The $x$-axis was obtained
by interpolating between two known minima, whereas the $y$-axis was chosen randomly.
One can readily verify from the figure that the optimization task is indeed extremely hard.

The calculation of the holonomy requires evaluating the ordered product in Eq.~(\ref{eq:U}).
The method used in the numerical algorithm is to simply write
the ordered product in a finite-difference approximation by considering
the connection components as being constant over a small difference in the parameters $\delta\gamma_i$, i.e.
\begin{equation}
U_\gamma\approx 
\exp\left( -\mathcal{A}_i({\gamma_n}) \delta \gamma^i_n \right)\cdots 
\exp\left(-\mathcal{A}_i(\gamma_1) \delta \gamma^i_1 \right).
\end{equation}
Throughout the study we used 200 discretization points per edge, i.e., $n=200\times(k+1)$.

\begin{figure}
\begin{picture}(130,90)
\footnotesize
\put(-50,5){\includegraphics[width=0.45\textwidth]{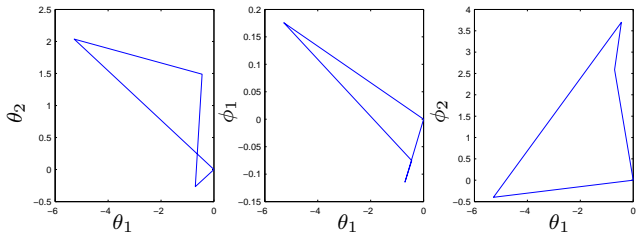}}
\put(145,0){$\theta_1$}
\put(65,0){$\theta_1$}
\put(-20,0){$\theta_1$}
\put(-60,40){\begin{sideways} $\theta_2$\end{sideways}}
\put(20,40){\begin{sideways} $\phi_1$\end{sideways}}
\put(100,40){\begin{sideways} $\phi_2$\end{sideways}}
\normalsize
\end{picture}
\caption{\label{fg:hada}Loop in parameter space that gives the Hadamard gate.}
\end{figure}

\begin{table}
  \begin{center}
    \begin{tabular}{|l|r|r|r|r|}
      \hline
    NODE& $\theta_1$ &$\theta_2$ & $\phi_1$ & $\phi_2$  \\ \hline
    begin & 0     &    0    &     0   &      0        \\ \hline
    1&    -5.28 &  2.04 &  0.18 & -0.40 \\ \hline
    2&   -0.44 &  1.49 &  -0.08 &   3.70 \\ \hline
    3&   -0.70 &  -0.27 & -0.11 &   2.59 \\ \hline
    end  &   0    &     0  &       0    &     0 \\ \hline
    \end{tabular}
\caption{\label{tab1} % data from pol3
\footnotesize{Loop of Fig.~\ref{fg:hada} numerically.}}
  \end{center}
\end{table}

\begin{figure}
\begin{picture}(130,90)
\footnotesize
\put(-50,5){\includegraphics[width=0.45\textwidth]{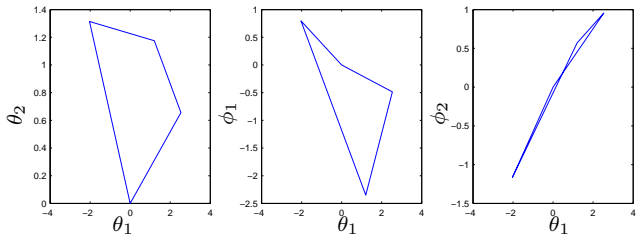}}
\put(145,0){$\theta_1$}
\put(65,0){$\theta_1$}
\put(-20,0){$\theta_1$}
\put(-60,40){\begin{sideways} $\theta_2$\end{sideways}}
\put(20,40){\begin{sideways} $\phi_1$\end{sideways}}
\put(100,40){\begin{sideways} $\phi_2$\end{sideways}}
\normalsize
\end{picture}
\caption{\label{fg:example}Loop in parameter space that yields the gate 
$U=e^{i}\exp\left(i\frac{\pi}{7}\sigma_z\right)\exp\left(i\frac{1}{3}\sigma_y\right)\exp{i\sigma_z}$.}
\end{figure}

\begin{table}
  \begin{center}
    \begin{tabular}{|l|r|r|r|r|}
      \hline
    NODE& $\theta_1$ &$\theta_2$ & $\phi_1$ & $\phi_2$  \\ \hline
    begin & 0     &    0    &     0   &      0        \\ \hline
    1&   -2.03 &   1.31 &   0.80 & -1.16 \\ \hline
    2&      1.21 &   1.18 &  -2.35 &   0.57 \\ \hline
    3&   2.54 &   0.66 &  -0.49 &   0.96 \\ \hline
    end  &   0    &     0  &       0    &     0 \\ \hline
    \end{tabular}
\caption{\label{tab2} % data from pol3
\footnotesize{Loop of Fig.~\ref{fg:example} numerically.}}
  \end{center}
\end{table}

\begin{figure}
\begin{picture}(250,345)
\put(-10,-10){\includegraphics[width=0.51\textwidth]{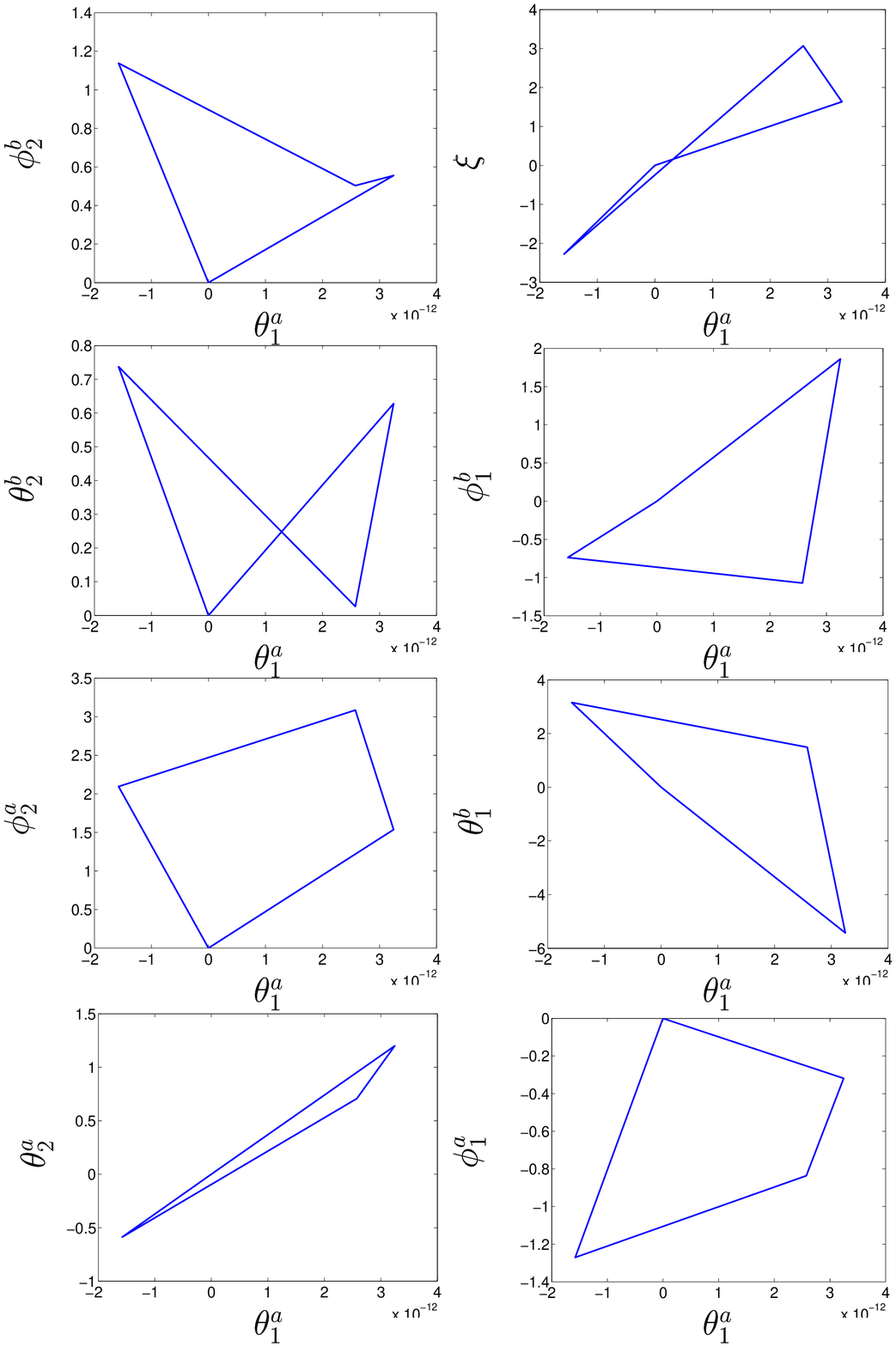}}
\end{picture}
\caption{\label{fg:CNOT}Loop in parameter space that gives the CNOT gate. Here $\gamma_\mathrm{CNOT}\in\mathcal{V}_3$ and
the error is below $10^{-13}$.}
\end{figure}

\begin{figure}
\begin{picture}(250,345)
\put(-10,-10){\includegraphics[width=0.51\textwidth]{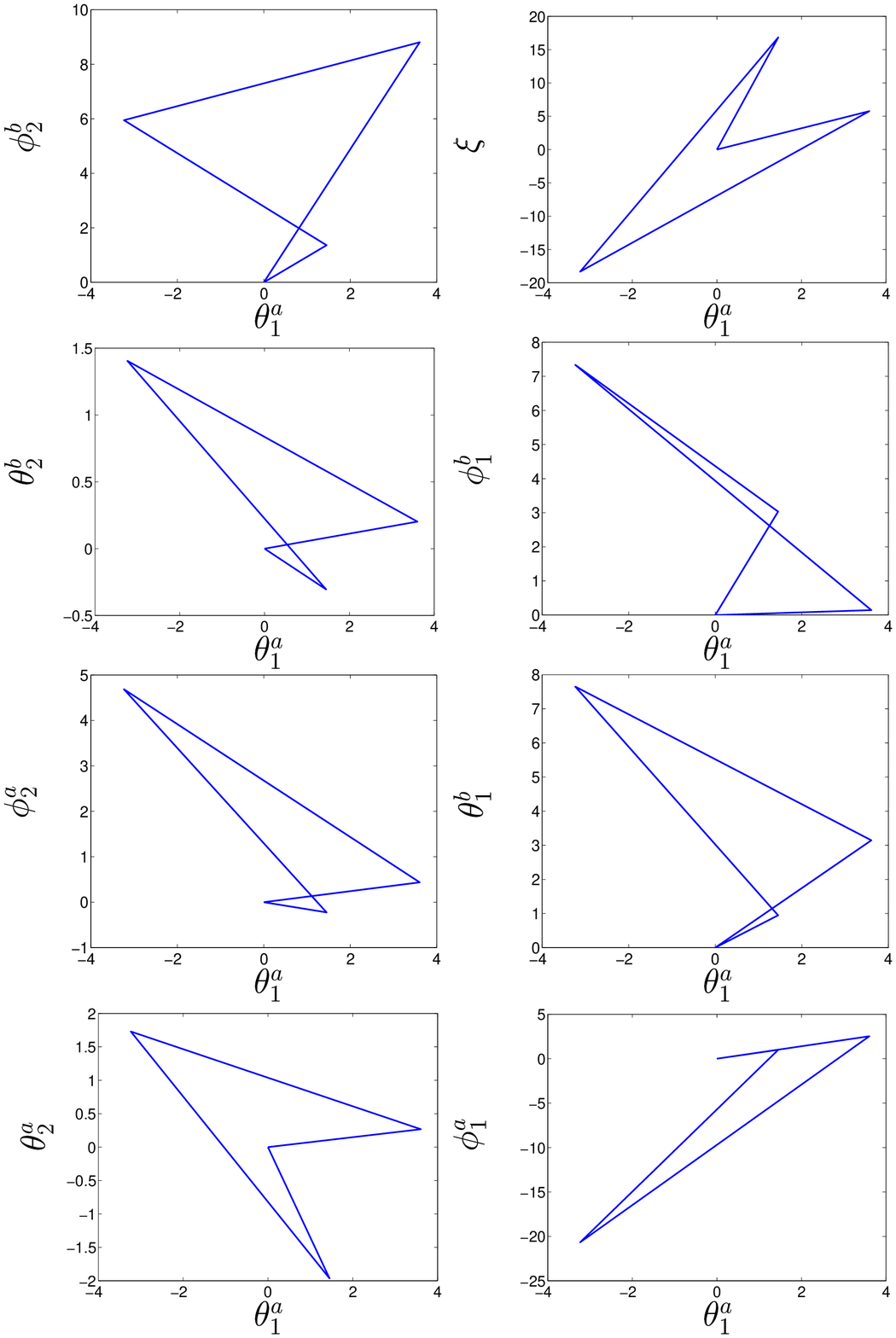}}
\end{picture}
\caption{\label{fg:swap}Loop in parameter space which realizes
the SWAP gate. 
Here the error is below $10^{-13}$.  In this case the variational space is $\mathcal{V}_5$.}
\end{figure}

\begin{figure}
\begin{picture}(250,345)
\put(-10,-10){\includegraphics[width=0.51\textwidth]{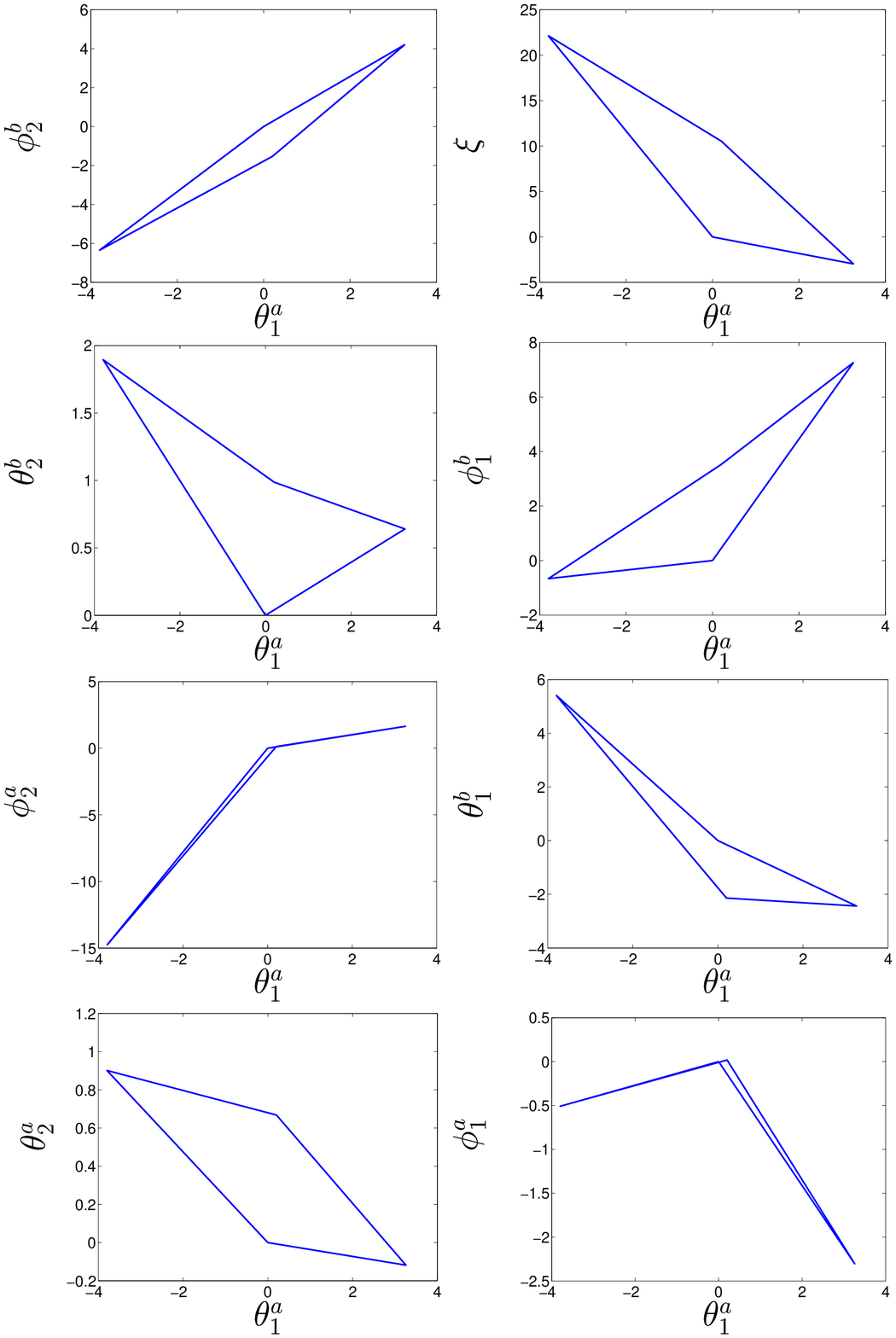}}
\end{picture}
\caption{\label{fg:fourier} Loop $\gamma_\mathrm{Fourier}$.
The error is
below $10^{-13}$.}
\end{figure}

\section{Results}\label{sec5}
First we attempted to find a loop that yields the Hadamard gate.Using a random initial configuration, 
we obtained the results that are plotted in Fig.~\ref{fg:hada}. The error function $f(\gamma)$ had a value smaller than $10^{-8}$ 
at the numerical optimum.The plot represents all the 
possible projections on two perpendicular axes (the horizontal axis is always $\theta_1$) in the four-dimensional space. 
Note that this optimization was carried
out in $\mathcal{V}_3$, meaning that there are three vertices other than the reference point. The results do not take advantage of
the $2\pi$ periodicity. 
We have also included the data points in Table~\ref{tab1}. It is impressive that such a simple control loop
yields the gate. Furthermore, this is just one implementation of the Hadamard gate. It is possible to find many different ones.

Another example of one-qubit gates is given in Fig.~\ref{fg:example} and in Table~\ref{tab2}. The gate that we tried to implement
was now chosen arbitrarily to be $U=e^{i}\exp\left(i\frac{\pi}{7}\sigma_z\right)\exp\left(i\frac{1}{3}\sigma_y\right)\exp{i\sigma_z}$.
Again, the error was well below $10^{-8}$ at the optimum. We argue that our method is capable of finding
any one-qubit gate. These results are not very enlightening as such, but should nevertheless clearly prove the strength of the technique. 

We also found several implementations for two-qubit gates.
Figure~\ref{fg:CNOT} presents the loop $\gamma_\mathrm{CNOT}\in\mathcal{V}_3$ that produces the CNOT. 
We observe, however, that again the minimization
resulted in an accurate solution. The minimization landscape is
just as rough in the case of two qubits. Now, of course, the dimension of $\mathcal{V}_3$ is
24. 

We also found an implementation of the SWAP gate given in Fig.~\ref{fg:swap}.

Finally, it is interesting to observe that even the two-qubit quantum Fourier transform can
be performed easily.
The resulting loop
is presented in Fig.~\ref{fg:fourier}.
It is remarkable that such a simple single loop yields a two-qubit quantum Fourier transform.
We used only three vertices but were still able to find an acceptable solution.
We argue that the error can be made arbitrarily small for any two-qubit gate.

\section{Discussion}\label{sec6}

The realization of arbitrary one- and two-qubit gates 
in the context of holonomic quantum computation has been
demonstrated. By restricting the loops in the control manifold within
a polygon with $k$ vertices, it becomes possible to cast the
realization problem to a finite-dimensional variational problem.
We have shown explicitly that some useful two-qubit gates are
realized by a single loop.

A possible improvement of the present scenario would be to  minimize the
length of the path realizing a given gate. This can be carried out
by introducing an appropriate penalty or barrier function and the Fubini-Study
metric in the control manifold $\mathbb{C}P^2$. This 
optimization program is under progress and will be reported elsewhere.

\begin{acknowledgments}

AN would like to thank the Research Foundation of Helsinki University of Technology 
and the Graduate School in Technical Physics for financial support; 
MN thanks the Helsinki University of Technology for a Visiting Professorship and he is also grateful for partial support of   
Grant-in-Aid from the Ministry of Education, Science, Sports and Culture, Japan (Project Nos. 14540346 and 13135215); 
MMS acknowledges the Academy of Finland for a Research Grant in Theoretical Materials Physics.

\end{acknowledgments}

%\newpage %Just because of unusual number of tables stacked at end

\begin{thebibliography}{99}

\bibitem{gruska}
J. Gruska,
\emph{Quantum Computing},
McGraw-Hill, New York (1999).

\bibitem{book}
 M. A. Nielsen and I.L. Chuang,
\emph{Quantum Computation and Quantum Information},
Cambridge University Press, Cambridge (2000).

\bibitem{review}
A. Galindo and M. A. Martin-Delgado, 
%\emph{Information and computation: Classical and quantum aspects}, 
Rev. Mod. Phys. {\bf 74}, 347(2002).

\bibitem{holonomic}
P. Zanardi and M. Rasetti,
%\emph{Holonomic quantum computation},
Phys. Lett. A. {\bf 264}, 94 (1999).

\bibitem{fujii}
K. Fujii,
%\emph{Mathematical foundations of holonomic quantum computer},
Rep. Math. Phys. {\bf 48}, 75 (2001).

\bibitem{univ}
D. Ellinas and J. Pachos,
%\emph{Universal quantum computation by holonomic and nonlocal gates with imperfections},
Phys. Rev. A. {\bf 64}, 022310 (2001).

\bibitem{nonabelian}
J. Pachos, P. Zanardi and M. Rasetti,
%\emph{Non-Abelian Berry connections for quantum computation},
Phys. Rev. A. {\bf 61}, 010305(R) (1999).

\bibitem{holonomies}
J. Pachos, P. Zanardi,
%\emph{Quantum holonomies for quantum computing},
Int. J. Mod. Phys {\bf 15}, 1257 (2001).

\bibitem{optical}
J. Pachos and S. Chountasis,
%\emph{Optical holonomic quantum computer},
Phys. Rev. A. {\bf 62}, 052318 (2000).

\bibitem{berry}
M. Berry,
%\emph{Quantal phase factors accompanying adiabatic changes},
Proc. R. Soc. Lond. A {\bf 392}, 45 (1984).

\bibitem{zee}
A. Zee,
%\emph{Non-Abelian gauge structure in nuclear quadrupole resonance},
Phys. Rev. A {\bf 38}, 1 (1988).

\bibitem{wilczek-zee}
F. Wilczek and A. Zee,
%\emph{Appearance of gauge structure in Simple Dynamical Systems},
Phys. Rev. Lett. {\bf 52}, 2111 (1984).

\bibitem{nakahara}
M. Nakahara,
\emph{Geometry, Topology and Physics},
IOP Publishing Ltd., Bristol (1990).


\bibitem{poly}
J. C. Lagarias, J. A. Reeds, M. H. Wright, and P. E. Wright, 
\emph{Convergence Properties of
the Nelder-Mead Simplex Method in Low Dimensions},
 SIAM Journal of Optimization {\bf 9}, 112 (1998).
\end{thebibliography}

\end{document}